\documentclass[twocolumn,showpacs,preprintnumbers,amsmath,amssymb]{revtex4}

\usepackage{graphicx}
\usepackage{dcolumn}
\usepackage{bm}

\begin{document}

\title{Charge inversion at minute electrolyte concentrations}

\author{J. Pittler,$^{1}$ W. Bu,$^{2}$ D. Vaknin,$^{2}$ A. Travesset,$^{2}$ D. J. McGillivray,$^{3}$ and M. L\"{o}sche$^{3}$}
\affiliation{$^{1}$Institute of Experimental Physics I, University of Leipzig, D-04103 Leipzig, Germany;\\
$^{2}$Ames Laboratory and Department of Physics and Astronomy, Iowa State University, Ames, Iowa 50011;\\
$^{3}$CNBT Consortium, NIST Center for Neutron Research, Gaithersburg, MD 20899\\
and Department of Physics, Carnegie Mellon University, Pittsburgh, PA 15213}
\date{\today}

\begin{abstract}

Anionic DMPA monolayers spread on LaCl$_3$ solutions reveal strong cation adsorption and a sharp transition to surface overcharging at unexpectedly low bulk salt concentrations. We determine the surface accumulation of La$^{3+}$ with anomalous x-ray reflectivity and find that La$^{3+}$ compensates the lipid surface charge by forming a Stern layer with $\approx 1$ La$^{3+}$ ion per 3 lipids below a critical bulk concentration, $c_t \approx 500\,\mathrm{nM}$. Above $c_t$, the surface concentration of La$^{3+}$ increases to a saturation level with $\approx 1$ La$^{3+}$ per lipid, thus implying that the total electric charge of the La$^{3+}$ exceeds the surface charge. This overcharge is observed at $\approx$ 4 orders of magnitude lower concentration than predicted in ion-ion correlation theories. We suggest that transverse electrostatic correlations between mobile ions and surface charges (interfacial Bjerrum pairing) may account for the charge inversion observed in this dilute regime.
\end{abstract}
\maketitle

Counterion screening of charged interfaces in electrolytic solutions is key for understanding a broad range of phenomena in molecular biology, colloid and polymer science, or microfluidics. Directly or indirectly, charges at cell membrane surfaces control functions and conformations of nearby molecules that may be involved in inter- and intracellular transport processes, cell-cell recognition and biomimetic mineralization processes. Despite extensive experimental and theoretical work performed over more than a century, problems regarding the nature of ion correlations, the role of the hydration sheath in physical processes and the structure of water near charged interfaces remain the subject of intense theoretical and experimental work \cite{Boroud05,Grosberg2002,Levin2002}.

The phenomenology of the electrostatics in soft media is rich and complex. Effects such as {\em charge inversion}, where charges at an interface attract counterions in excess of their own nominal charge density have been reported in the literature \cite{McLaughlin1989,Boroud05,Besteman2004,Besteman2005,Grosberg2002}. Current theories assume that charge inversion results from the free energy gain brought about by ion-ion correlations \cite{Grosberg2002} and predict that measurable charge inversion occurs when $q /(\pi \sigma \lambda_D^2)\sim 1$, where $q = Z e$ is the counter-ion charge and $\sigma$ is the surface charge density. $\lambda_D = (8\pi l_B I)^{-1/2}$ is the Debye length of an electrolytic solution (Bjerrum length, $l_B = e^2/(\epsilon k_BT)$) with ionic strength $I$ \cite{Safran1994}. Applying this reasoning to Langmuir films of a charged lipid such as phosphatidic acid, one expects for a typical area per amphiphile, $A_{\mathit{lipid}} (= e/\sigma$) $\approx$ 40 \AA$^2$, spread on LaCl$_3$ solution ($Z=3$), that the salt concentration has to be at least in the 10 mM range for charge inversion to become significant. Besides correlations of the mobile ions, there are other effects that may contribute to the phenomenon. For example, it has recently been suggested that {\em transverse} correlations, {\em i.e.}, correlations between interfacial charges and mobile ions, may also play a role in the generation of charge inversion \cite{Henle04,Travesset06}.

In this paper, we demonstrate charge inversion at  ultra-low ($\mu \mathrm{M}$ or less) LaCl$_3$ electrolyte concentrations with surface-sensitive, resonant X-ray scattering at the La L$_{III}$ absorption edge and discuss its relationship to proton transfer and release.  These results extend previous studies of BaCl$_2$ \cite{Vaknin2003,Schalke2000_1} and CsI solutions \cite{Bu2005,Bu2006}.


Surface monolayers of DMPA (1,2-dimyristoyl-\textit{sn}-glycero-3-phosphatidic acid) (Avanti Polar Lipids), were spread from chloroform/methanol (3:1; both from Merck, \textit{p.a.} grade) on aqueous solutions with different LaCl$_3$ (Sigma, purity: $>$ 99\%) solutions prepared from ultrapure water (NANOpure, Barnstead) in an enclosed Teflon (PTFE) Langmuir trough. For the x-ray scattering work, the air space above the trough was continuously purged with water-saturated helium. Both isotherm and scattering studies were performed at a temperature of 21$^{\circ}\mathrm{C}$. To minimize ion contamination, the subphases were handled and transferred to the Langmuir film balance in bottles made from Teflon (Fisher Scientific). After solvent evaporation, the monolayers were compressed at a rate of $\sim 1\,$\AA$^2 / (\text{molecule min.})$ to record isotherms or measure reflectivities at various surface pressures between $\pi = 10$ and $50\,\mathrm{mN/m}$. At bulk ion concentrations in the $\mu$M range, monolayers become progressively unstable to compression. At a collapse pressure, $\pi_c$, the compressibility of the surface film increases, consistent with a gradual transformation of the monolayer locally into bilayer or trilayer structures. $\pi_c$ decreases with increasing salt concentration, which makes measurements at high $\pi$ difficult for $c_{\mathit{bulk}}^{LaCl_3} > 1\,\mu\mathrm{M}$. We restrict the following discussion to the low surface pressure regime, $\pi = 15 \ldots 30\,\mathrm{mN/m}$.

X-ray reflectivity (XR) measurements were conducted on the Ames Laboratory horizontal surface diffractometer at beamline 6-ID-B of the Advanced Photon Source (APS). The highly monochromatic beam (energy resolution, $\Delta E \sim 1\,\mathrm{eV}$), selected by an initial Si double crystal monochromator, is deflected onto the liquid surface at a specified incidence angle by a secondary Ge(111) monochromator located at the diffractometer \cite{Vaknin2001}. The x-ray energy $E$ was calibrated by measuring the absorption spectrum of a dilute LaCl$_3$ solution. Off-resonance spectra were measured at $8.0\,\mathrm{keV}$, and ``on-resonance'' refers to the La $L_{III}$ edge at $E_{\mathit{res}} = 5.486\,\mathrm{keV}$. To reduce radiation damage, the Langmuir trough was periodically translated across the beam. To monitor sensitively for damage, in particular after prolonged exposure of the sample at high momentum transfer, $Q_z$, measurements were routinely repeated -- without translating the sample -- across the $Q_z$ regions of the sharp cusps in the interference minima.

To extract scattering length density (SLD) distributions  across the interface from the XR data, parameterized profiles $\rho(z) = \rho^{\prime}(z)+i\rho^{\prime\prime}(z)$ were constructed in which the real and imaginary parts of $\rho$ describe the electron density (ED) and the absorption density (AD), respectively, along the surface normal, $z$. We used both a modified 'box' model approach \cite{AlsNielsen1989,Bu2006} and the quasi-molecular \textit{Volume-Restricted Distribution Function} (VRDF) approach \cite{Schalke2000_1,Schalke2000_2} for this reconstruction to obtain stable results and to check for consistency. In the box model, $\rho(z)$ is described by a sum of error functions
\begin{equation}
\rho(z)=\frac{1}{2}\sum^{N}_{j=1}\mbox{Erf}\left(\frac{z-z_{j}} {\sqrt{2}\sigma_{j}}\right)(\rho_{j}-\rho_{j+1})+\frac{\rho_{N+1}}{2}
\label{Erf}
\end{equation}
where $N$ is the total number of slabs $j$ that describe the monolayer. $z_{j}$ and $\sigma_{j}$ denote the position and r.m.s. roughness of the $j$th interface, respectively, between the slabs, and $j = N+1$ indicates the bulk subphase. Alternatively, $\rho(z)$ is constructed from a sequence of thermally broadened distributions, modeled as Gaussian functions, that represent parsed subfragments of the lipid structure and bound cations \cite{Schalke2000_1,Vaknin2003}. In both approaches, volume filling is implemented by accounting for the subfragment volumes as derived from molecular dynamics simulations \cite{Armen1998}. For either model, $\rho^{\prime\prime}(z)$ is only significant in spectra measured at $E_{\mathit{res}}$ and correlates with the physical distribution of the cations across the interface, \textit{i.e.}, it is negligible where the cation density is at bulk density or below. Reflectivities are calculated from $\rho(z)$ by applying a recursive method \cite{Parratt1954} to the discretized profiles. Both $\int{\Delta \rho^{\prime}(z)~dz} = \int{\left(\rho_{\mathit{off-res}}^{\prime}(z)- \rho_{\mathit{res}}^{\prime}(z)\right)~dz}$, and -- more directly -- $\int{\rho^{\prime\prime}(z)~dz}$ provide a direct and model-independent measure of the number of La$^{3+}$ adsorbed at the interface:
\begin{equation}
n_{La^{3+}}=\frac{{\int{\Delta \rho^{\prime}~dz}}}{\Delta z_{\mathit{eff}} \cdot A_{\mathit{lipid}}}
\label{n_La}
\end{equation}
where $\Delta z_{\mathit{eff}} \sim 19.0$ is the difference between the effective numbers of electrons for La$^{3+}$ at $E_{\mathit{off-res}}$ and $E_{\mathit{res}}$.

\begin{figure}[htl]
\includegraphics[width=6 cm]{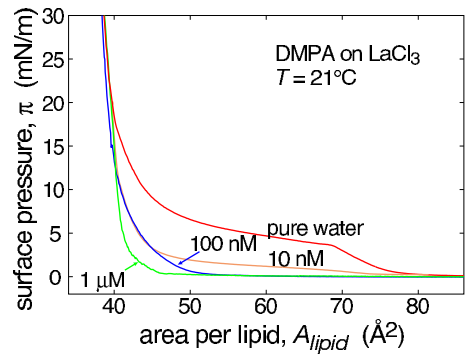}
\caption{DMPA monolayer isotherms for different La$^{3+}$ subphase concentrations} \label{iso} \end{figure}

\begin{figure}[htl]
\includegraphics[height=8.7cm]{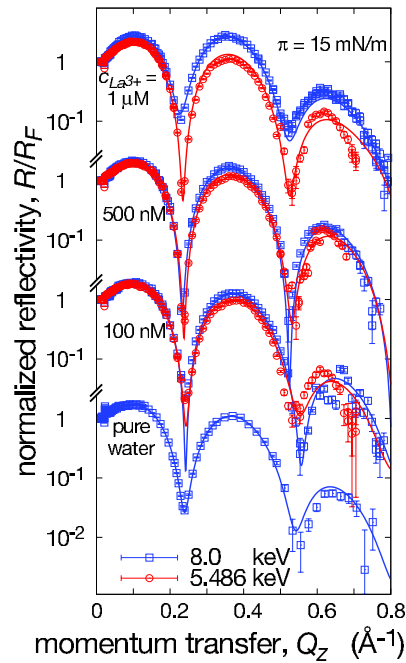} \caption{Fresnel-normalized x-ray reflectivities, $R/R_F$, of DMPA monolayers at $E = 8.0\,\mathrm{keV}$ (off-resonance) and $5.486\,\mathrm{keV}$ (La $L_{III}$ resonance) on aqueous LaCl${_3}$ solutions. For clarity, subsequent pairs of data sets are offset by a factor of 100 each. Solid lines derive from model fits.} \label{ref1}
\end{figure}

\begin{figure}[htl]
\includegraphics[height=7.8cm]{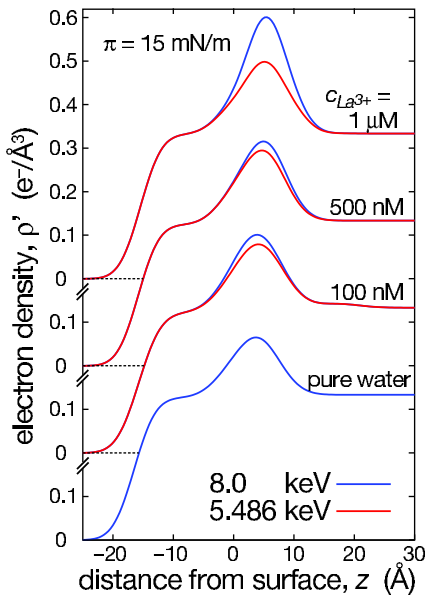}
\caption{VRDF electron density profiles derived from fits to the experimental data. The off-resonance and on-resonance data shown in Fig.~\ref{ref1} were co-refined with consistent sets of parameters that deviated only in the effective electron numbers and absorption cross sections of the La$^{3+}$ component. Subsequent ED pairs are offset by $0.2\,e^{-}$/\AA$^{3}$ for clarity. Similar ED profiles were derived from a modified box models as described in the text. The number density of La$^{3+}$ at the interface was subsequently obtained from such models via Eq.~\ref{n_La}.} \label{ref2}
\end{figure}

\begin{figure}[htl]
\includegraphics[width=5.5cm]{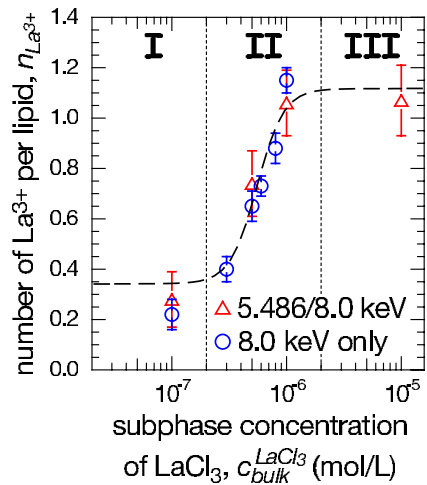}
\caption{Number of adsorbed La$^{3+}$ per DMPA in surface monolayers as a function of bulk LaCl$_3$ concentration.
Red plot symbols indicate results at 15 mN/m derived from the anomalous reflectivity data pairs shown in Fig.~\ref{ref1} and ED profiles similar to those shown in Fig.~\ref{ref2}. The results indicated by blue symbols were derived from independent data measured at $8.0\,\mathrm{keV}$ alone.} \label{La-DMPA}
\end{figure}


Figure~\ref{iso} shows isotherms, $\pi$ \textit{vs.} $A_{\mathit{lipid}}$,  of DMPA monolayers at various LaCl$_3$ concentrations of the bulk subphase. On the salt-free subphase, a sloped plateau indicates coexistence of a liquid and a hexatic lipid phase. The height of this plateau depends sensitively on salt concentration. As shown earlier, micromolar concentrations of Ca$^{2+}$ condense the monolayer to the extent that this plateau disappears \cite{Loesche1989}. Figure~\ref{iso} shows that trivalent cations are even more effective in monolayer condensation and the plateau has already been lost at $c_{\mathit{bulk}}^{LaCl_3} > 10\,$\textit{nano}M, consistent with strong electrostatic interactions between La$^{3+}$ and the DMPA${^-}$ headgroups. It has been shown with fluorescence microscopy that Ca$^{2+}$ condenses DMPA$^-$ from the two-dimensional (2D) gas phase into hexatic domains \cite{Loesche1989}. We expect that binding of La$^{3+}$ to DMPA$^-$ has a similar effect already at exceedingly low salt concentrations.

Figure~\ref{ref1} shows exemplary, Fresnel-normalized off-resonance/on-resonance reflectivity pairs of DMPA monolayers at $\pi = 15\,\mathrm{mN/m}$ on subphases that contain 100, 500, and $1\,\mu\mathrm{M}$ LaCl$_3$. A high-energy reflectivity of DMPA on pure water is also shown. Qualitatively, the increase in intensity with increasing LaCl${_3}$ bulk concentration indicates ED increases due to the accumulation of La$^{3+}$ at the acidic interfaces. Quantitatively, the difference between off-resonance and on-resonance EDs is proportional to the La$^{3+}$ surface concentration (Eq.~\ref{n_La}). The data fall in two categories: Between  $c_{\mathit{bulk}}^{LaCl_3} = 0$ and $500\,\mathrm{nM}$ LaCl$_3$, both off-resonance and resonance reflectivities are rather similar. Above $500\,\mathrm{nM}$ LaCl$_3$, on the other hand, the off-resonance reflectivities are significantly higher at all $Q_z$ than those at resonance. Without any model interpretation, this qualitative difference shows conclusively that the condensation of La$^{3+}$ at the monolayer is significantly larger at $1\,\mu\mathrm{M}$ LaCl$_3$ than at $500\,\mathrm{nM}$ and below. This is corroborated and quantified by a detailed data analysis (see Fig.~\ref{ref2}), using the modified box model and the VRDF. We also measured more complete, systematic sets of off-resonance reflectivities over a wider range of subphase salt concentrations, between $100\,\mathrm{nM}$ and $10\,\mu\mathrm{M}$, and at higher surface pressures. As shown in Fig.~\ref{La-DMPA}, these results are consistent with the more stringent anomalous reflectivity data.

Figure~\ref{ref2} shows SLD profiles $\rho(z)$ obtained from a co-refinement of data pairs shown in Fig.~\ref{ref1}. While $\rho^{\prime}(z)$ in the lipid headgroup region is larger for the off-resonance data than for the on-resonance data on all LaCl$_3$-containing subphases, the difference is much larger for $1\,\mu\mathrm{M}$ LaCl$_3$. The VRDF modeling suggests that the differences in $\rho^{\prime}$, to be interpreted as the distribution of cations at the interface, are confined to the lipid headgroup region.  Equation~\ref{n_La} serves to determine $n_{\text{La}^{3+}}$, the number of La$^{3+}$ per DMPA headgroup in the monolayer. A compilation of this data collected at various surface pressures as a function of  $c_{\mathit{bulk}}^{LaCl_3}$, is given in Fig.~\ref{La-DMPA}. The plot shows clearly three distinct regimes which consist of two plateaus at low and high $c_{\mathit{bulk}}^{LaCl_3}$, separated by a sharp transition. In regime I, between $100\,\mathrm{nM}$ and $\approx 200\,\mathrm{nM}$ LaCl$_3$, $n_{\text{La}^{3+}}$ is constant at $0.34 \pm 0.1$. It is followed by a remarkably sharp transition, regime II, that connects this regime with regime III, observed at high LaCl$_3$ concentration ($> 1\,\mu\mathrm{M}$), where $n_{\text{La}^{3+}} \approx 1$.

The results in regime I correspond to $\approx 1$ La$^{3+}$ bound to three DMPA at the interface. This stoichiometry follows from the dissociation constants of the first and second protons to DMPA, pK$_{a,1} = 2.1$ and pK$_{a,2} = 7.1$ \cite{Atkins}, and the binding constant of La$^{3+}$ to a monovalent phosphate group, estimated as $K_B \approx 100\,\mathrm{M}^{-1}$ \cite{Travesset06}, introduced into the PB theory \cite{Israelachvili2000,Travesset06}. Regime II follows from the large affinity of La$^{3+}$ ions to divalent phosphate groups at the interface, $K_B \sim 10^{5}-10^{6}\,\mathrm{M}^{-1}$ \cite{Travesset06}, which leads to deprotonation, PO$_4$H$^- \rightarrow$ PO$_4^{^{2-}}$, of the interfacial group at a concentration $c^{La^{3+}}_{\mathit{bulk}} \sim 10^{-6}$. The increase of $n_{\text{La}^{3+}}$ from 1/3 to 2/3 is therefore not due to charge inversion, but rather to a doubling of interfacial charge density by surface deprotonation. Regime III with $n_{\text{La}^{3+}}\sim 1$, on the other hand, clearly shows charge inversion that is not covered by established ion-ion correlation theories \cite{Grosberg2002}. The experimental observation of charge inversion in this concentration range is thus compelling evidence for yet other effects of the interface \cite{Travesset06}. A hypothetical Stern layer, that would consist of La$^{3+}$ bound to interfacial PO$_4^{2-}$ in a 1:1 ratio ($n_{La^{3+}} \approx 1$), leads inevitably to a diffuse distribution of co-ions (\textit{e.g.}, Cl$^-$) which results in a free energy penalty (per amphiphile) of \cite{Safran1994}
\begin{equation}\label{PB_free}
F\approx 2(\ln(2\sqrt{2}\lambda_D/\lambda_{GC})-1)\approx 18 k_B T,
\end{equation}
where $\lambda_{GC} = e/(2\pi \sigma l_B)$ is the Gouy-Chapman length.  We envisage two scenarios, under which such a free energy penalty could be compensated by other effects: (1) A ``molten salt'' state consisting La$^{3+}$ ions intercalated between the negative charges on the phosphate oxygens forming a highly correlated state \cite{Travesset06}. (2) Enhanced adsorption in the Stern layer via non-electrostatic interactions, for example, hydrogen bonding between the PO$_4^{2-}$ oxygens (acceptor) and a La(OH)$^{2+}$ complex (donor). In either scenario, charge inversion is associated with some instability of the Langmuir monolayer, as evidenced by slow collapse at intermediate surface pressures at LaCl$_3$ concentrations above $5\cdot 10^{-5}\,\mathrm{M}$.


In this paper we presented experimental evidence for charge inversion at minute electrolyte concentrations ($0.5\,\mu\mathrm{M}$ LaCl$_3$) in contact with acidic DMPA monolayers. When compared with experimental work investigating forces between charged plates \cite{Besteman2005}, charge inversion in our experiment is observed at $\approx 2$ orders of magnitude lower and in the absence of any monovalent salt. More fundamentally, our results provide strong indication that correlations among mobile ions and interfacial charges play a prominent role in the generation of charge inversion. We interpret the concentration dependence of counter-ion accumulation beyond regime (I) in Fig. \ref{La-DMPA} as follows: Charge inversion is initiated by the deprotonation of the secondary oxygen in the phosphate group (regime II). The high level of charge inversion at $\mu\mathrm{M }$ bulk LaCl$_3$ (regime III) requires additional effects of ion reorganization, and we suggested two tentative scenarios.

The results presented in this study corroborate the notion that charge inversion is a generic effect for multivalent ions next to highly charged surfaces and quantify the role of multivalent ions in favoring surface deprotonation. Furthermore, however, these results are likely of universal relevance for the physics of highly charged colloidal or biological systems in the presence of multivalent ions, as evidenced from previous studies \cite{McLaughlin1989,Vaknin2003}. Finally, it is also clear that still more experimental and theoretical work is needed to fully resolve the fundamental issues involved in the electrostatics of highly charged interfaces.


We thank D. S. Robinson and D. Wermeille for technical support at the 6-ID beamline, and to J. Faraudo for helpful discussions. The MUCAT sector at the APS is supported by the U.S. DOE, Basic Energy Sciences, Office of Science, through Ames Laboratory under contract No. W-7405-Eng-82. Use of the Advanced Photon Source is supported by the U.S. DOE, Basic Energy Sciences, Office of Science, under Contract No. W-31-109-Eng-38. ML was supported by the NSF (grant no. 0555201), the NIH (grant no. 1 RO1 RR14812), the Regents of the University of California, and the Volkswagen Foundation (grant no. I/77709). AT is partially supported by NSF-DMR-0426597.


\end{document}